# Atomic Layer MoS$_2$-Graphene van der Waals Heterostructure Nanomechanical Resonators

Fan Ye, Jaesung Lee, Philip X.-L. Feng[*]

*Department of Electrical Engineering & Computer Science, Case School of Engineering, Case Western Reserve University, 10900 Euclid Avenue, Cleveland, OH 44106, USA*

## *Abstract*

**Heterostructures play significant roles in modern semiconductor devices and micro/nanosystems in a plethora of applications in electronics, optoelectronics, and transducers. While state-of-the-art heterostructures often involve stacks of crystalline epi-layers each down to a few nanometers thick, the intriguing limit would be heterto-atomic-layer structures. Here we report the first experimental demonstration of freestanding van der Waals heterostructures and their functional nanomechanical devices. By stacking single-layer (1L) MoS$_2$ on top of suspended single-, bi-, tri- and four-layer (1L to 4L) graphene sheets, we realize array of MoS$_2$-graphene heterostructures with varying thickness and size. These heterostructures all exhibit robust nanomechanical resonances in the very high frequency (VHF) band (up to ~100 MHz). We observe that fundamental-mode resonance frequencies of the heterostructure devices fall between the values of graphene and MoS$_2$ devices. Quality ($Q$) factors of heterostructure resonators are lower than those of graphene but comparable to those of MoS$_2$ devices, suggesting interface damping related to interlayer interactions in the van der Waals heterostructures. This study validates suspended atomic layer heterostructures as an effective device platform and opens opportunities for exploiting mechanically coupled effects and interlayer interactions in such devices.**

[*]Corresponding Author. Email: philip.feng@case.edu.

# Introduction

Heterostructures, often referring to stacks of thin films combining at least two materials with different band structures, are important building blocks in modern semiconductor devices and micro/nanosystems, especially in electronics, optoelectronics, and solid-state transducers. A basic working principle of heterostructures is to use 'bandgap engineering' for manipulating carriers, *e.g.*, electrons and photons at interfaces, by leveraging the offsets in bandgaps of different constitutive materials. In electronic domain, elementary components such as p-n junctions and various diodes (including light emitting diodes (LEDs) [1] and solar cells [2])are realized by stacking p-type and n-type thin layers. In optoelectronic and photonic domains, quantum wells and superlattices with periodically sandwiched thin layers are key to enabling many innovative diode lasers [3]. In mechanical domain, a classical example of heterostructures is a bimorph [4] that consists of two active layers and one passive layer, which is indispensable for many electromechanical actuators and sensors. To realize high-performance heterostructured devices, clean and abrupt interfaces are distinctly important, which have conventionally required highly advanced thin film growing and deposition techniques, such as molecular beam epitaxy (MBE) and atomic layer deposition (ALD). These have played significant roles and witnessed great successes and advances in creating state-of-the-art heterostructures in modern devices and technologies, attaining constitutive layers as thin as 9nm for selected materials (*e.g.*, Si/Ge, by MBE) [5].

The advent of atomically thin layered crystals and two-dimensional (2D) semiconductors (from graphene to transition metal dichalcogenides, *i.e.*, TMDCs) [6,7,8] offers new exciting opportunities for adding innovative members into the families of heterostructures, and new approaches to revolutionizing heterostructure devices, down to the ultimate limits of both discrete atomic layers and sharp hetero-interfaces. Enabled by atomic layer crystals, one can take a single-layer 2D material and stacking it on top of another single-layer 2D material. Distinct from traditional heterostructures that rely on strong chemical bonds between adjacent hetero-layers, such as ionic bonds and covalent bonds, the new atomically thin heterostructures are held together by van der Waal interactions [9]. Compared with devices made of individual 2D materials, van der Waals heterostructure devices demonstrate significant versatility and advantage in functions and performance, thus offering plentiful opportunities in both fundamental studies and device applications. For example, using atomic layers of h-BN as substrate, heterostructured graphene [10,11] and $MoS_2$ [12] FETs have been demonstrated with over tenfold mobility enhancement, with remarkable stability even under harsh conditions (high humidity and elevated temperature up to 500K) [13]. This mobility enhancement further allows the observation of Coulomb drag [14] and fractal quantum Hall effects [15], which has been elusive for devices with conventional $SiO_2$ insulator. In 2D heterostructure photovoltaics, the most studied are p-n junctions fabricated by combining p-type and n-type 2D semiconductor, *e.g.*, $WSe_2/MoS_2$ [16] junctions that show external quantum efficiency (EQE) from 10% to 30% with varying thickness [17]. In addition, phototransistors based on graphene/TMDCs/graphene structure exhibit extremely high EQE up to 30% [18]. Generally, two major approaches have been demonstrated for realizing van der Waals heterostructures: mechanical assembly including both solvent-assistant methods [10,19,20,21] and all-dry transfer techniques [22],epitaxy [16] or chemical vapor deposition (CVD) methods [23]. In comparison, mechanical assembly enables us to fabricate devices with high efficiency while CVD or epitaxy methods can achieve large scale heterostructures with finely controlled orientations between different layers.

Though important efforts have been made in van der Waals heterostructure devices, freestanding heterostructures of 2D crystals that possess mechanical degrees of freedom and controllable mechanical functions (such as bimorph) have not been explored yet.  In this work, we describe the first experimental demonstration of suspended heterostructures enabled by atomic layer crystals ($MoS_2$ on graphene) with varying thickness, and the first nanomechanical resonators based on these freestanding heterostructures.  We fabricate single-layer $MoS_2$ on single-layer graphene (1L$MoS_2$-1LGr), 1L$MoS_2$-2LGr, 1L$MoS_2$-3LGr and 1L$MoS_2$-4LGr (here Gr stands for graphene) heterostructure resonators, in circular drumhead geometry with various diameters.  We find all these heterostructure devices exhibit robust nanomechanical resonances, up to ~100MHz in the very-high frequency (VHF) band.  To systematically investigate resonance properties of the heterostructures, a set of control experiments of studying $MoS_2$ and graphene resonance properties are also performed.  We find that fundamental resonance frequencies of the heterostructures always fall between those of their $MoS_2$ and graphene counterparts.  In contrast, quality (*Q*) factors of heterostructure resonators are comparable to those of $MoS_2$ devices, but lower than those of graphene devices, implying an interlayer damping at the $MoS_2$-graphene interface.  Finally, tension levels of heterostructure resonators are quantified by matching experimental data with theoretical calculations.  These results provide evidence and initial insights for understanding interlayer interactions, and demonstrate a new platform for studying thermal properties and interlayer heat transfer of van der Waals heterostructures.

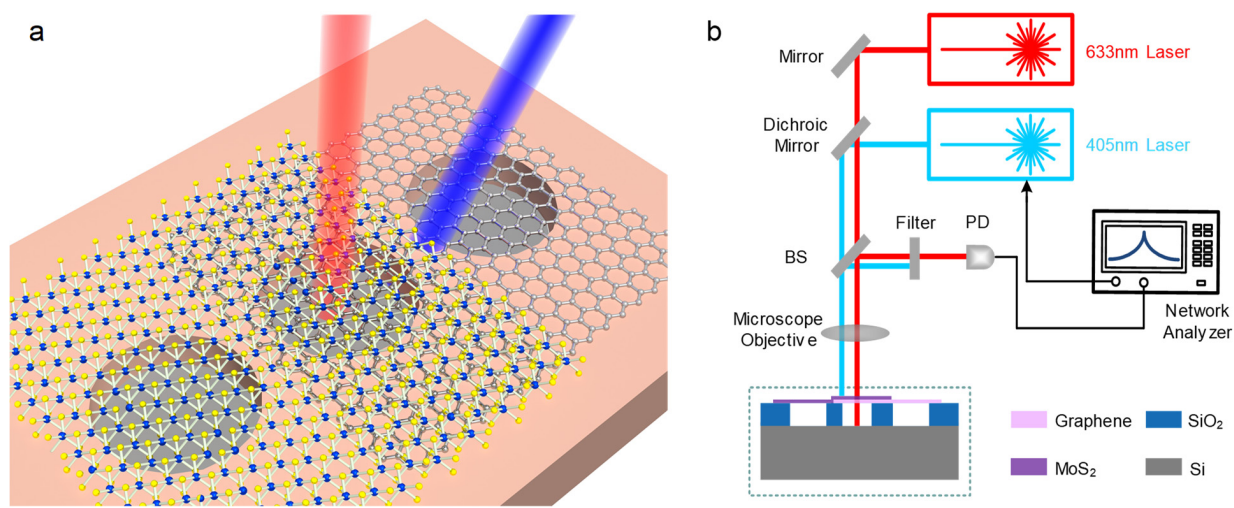

**Figure 1**: (a) Illustration of freestanding van der Waals heterostructures of $MoS_2$-graphene atomic layers.  Blue, yellow and silver spheres represent Mo, S and C atoms, respectively.  (b) Schematic of the nanomechanical resonance interferometry measurement system.  A 405nm blue laser is employed to excite the resonance motion, and a 633nm red laser is used for motion detection.  PD and BS represent photodetector and beam splitter, respectively.  All measurements are performed in moderate vacuum (~20mTorr) at room temperature.

## Results and Discussions

The fabrication of $MoS_2$-graphene heterostructure resonators starts from exfoliating graphene on $SiO_2$-on-Si substrates with pre-patterned arrays of microtrenches.  After obtaining atomically thin graphene, single-layer (1L) $MoS_2$ flakes are transferred on top of the existing graphene flakes using an all-dry transfer method with alignment [22].  CVD $MoS_2$ layers are used in 1L$MoS_2$-1LGr, 1L$MoS_2$-2LGr and 1L$MoS_2$-3LGr devices, while exfoliated $MoS_2$ flakes are used in making 1LMoS$_2$-4LGr devices. The 1LMoS$_2$-1LGr van der Waals heterostructure is illustrated in **Figure 1**a. To study how the nanomechanical resonance characteristics of heterostructure are affected by the constituting MoS$_2$ and graphene layers, only part of graphene sheet is covered by MoS$_2$, for clear control experiments. As shown in Fig. 1b, an intensity-modulated 405nm blue laser is employed to excite the resonance and a 633nm red laser is focused at the center of the suspended region of the device, to detect motion by exploiting ultrasensitive interferometry techniques. To reduce air damping, devices are preserved in vacuum during measurement.

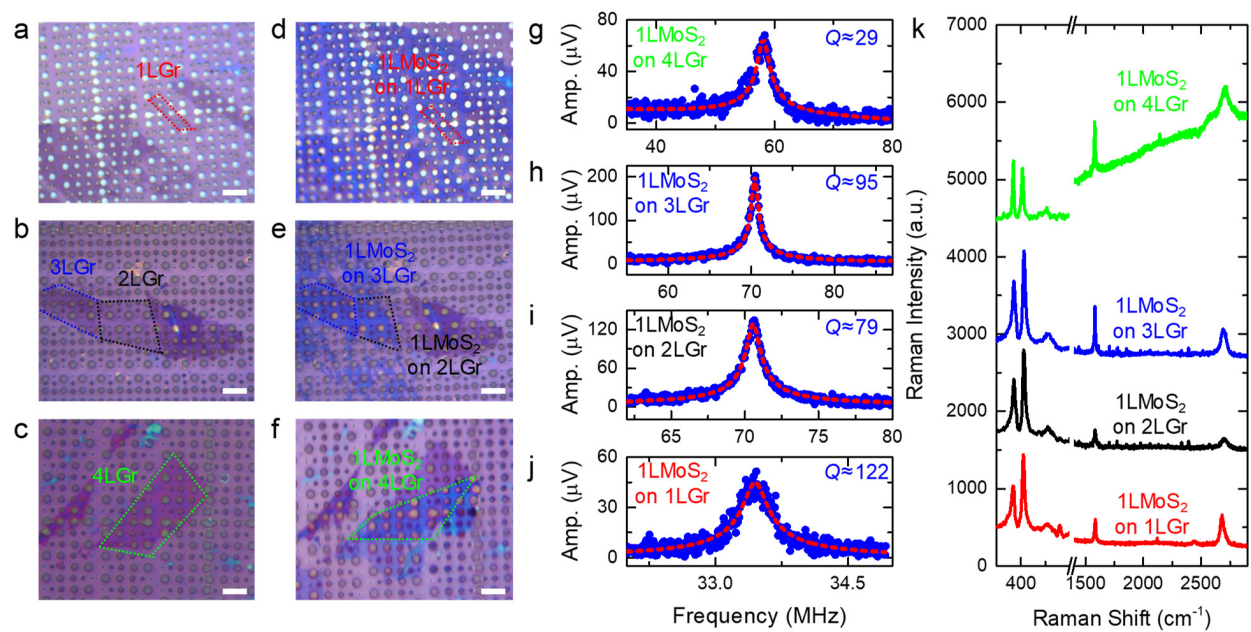

**Figure 2**: Microscopy images of (a) single-, (b) bi-, tri-, and (c) four-layer (1L to 4L) graphene flakes on substrates with arrays of circular microtrenches. Microscopy images of (d) 1LMoS$_2$ on 1LGr, (e) 1LMoS$_2$ on 2LGr, 1LMoS$_2$ on 3LGr, and (f) 1LMoS$_2$ on 4LGr suspended van der Waals heterostructures (here Gr stands for graphene) (Scale bar: 5µm). Fundamental-mode of one (g) 1LMoS$_2$ on 4LGr, (h) 1LMoS$_2$ on 3LGr, (i) 1LMoS$_2$ on 2LGr, and (j) 1LMoS$_2$ on 1LGr nanomechanical resonator. (k) Raman spectra of MoS$_2$-graphene heterostructure nanomechanical resonators.

Microscope images of 1L to 4L graphene nanosheets are shown in **Figure 2**a to c; and the corresponding heterostructure devices are illustrated in Fig. 2d to 2f. CVD MoS$_2$ flakes are used on top of 1L, 2L and 3L graphene while exfoliated MoS$_2$ is used on top of 4L graphene. As a result, we have attained totally 39 heterostructure devices, which facilitate us to study the statistics of the resonance characteristics. Figure 2g to 2j show the fundamental-mode resonances of devices of 1LMoS$_2$-4LGr, 1LMoS$_2$-3LGr, 1LMoS$_2$-2LGr and 1LMoS$_2$-1LGr, respectively. The measured resonance spectra are fitted to a damped simple harmonic resonator model to extract the fundamental-mode resonance frequency ($f_0$) and quality ($Q$) factor. The numbers of layers of the graphene and MoS$_2$ flakes are confirmed by Raman spectroscopy (Fig. 2k). The separation between $E^1_{2g}$ and $A_{1g}$ peaks in MoS$_2$ is around 18.5cm$^{-1}$ for exfoliated sample and 20.8cm$^{-1}$ for CVD sample, indicating that the MoS$_2$ flakes are 1L [24]. The number of graphene layers is identified by the ratio of peak intensities between G mode and 2D mode. In 1L graphene, G peak is lower than 2D peak. In 2L graphene, the intensities of G mode and 2D mode are almost the same. In 3L graphene, G peak is slightly stronger than 2D peak. The peak intensity difference between G mode and 2D mode becomes larger in 4L graphene [25].

Statistical results of resonance frequencies and *Q* factors of 1LMoS$_2$-2LGr, 1LMoS$_2$-3LGr, and 1LMoS$_2$-4LGr circular drumhead devices with different diameters are shown in **Table S1** in **Supplementary Information**. The fundamental-mode resonances of most of the devices are in the VHF band, with the highest frequency reaching ~100MHz. Interestingly, no clear size dependence of resonance frequency is observed in these devices, which could be attributed to different tension levels among these devices. The frequency versus *Q* factor plots are shown in **Fig. 3**a, 3b and 3c for 1LMoS$_2$-2LGr, 1LMoS$_2$-3LGr and 1LMoS$_2$-4LGr, respectively. The device with the highest figure-of-merit in each figure is highlighted by purple dash circle; the highest $f_0 \times Q$ obtained is $8.7 \times 10^9$ Hz in a 1LMoS$_2$-2LGr device.

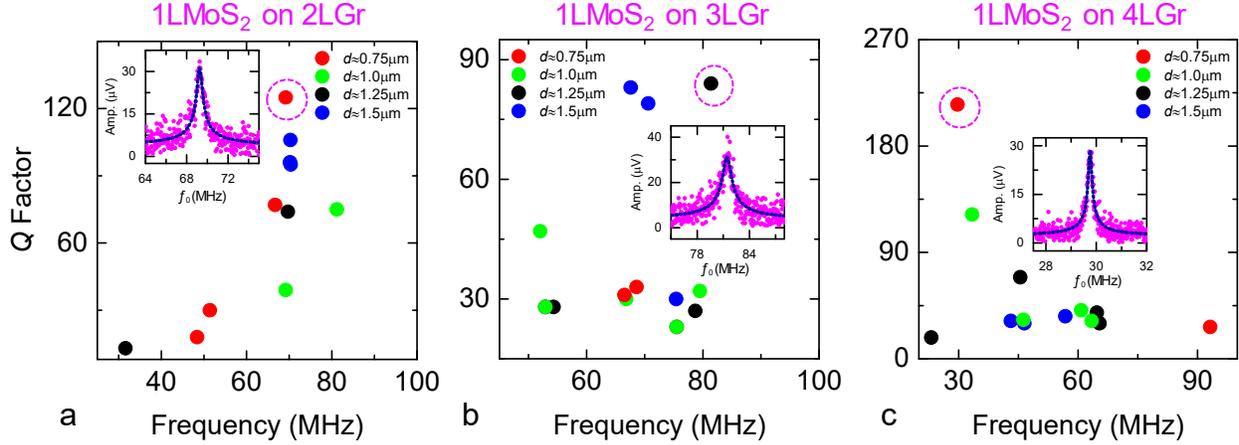

**Figure 3**: Fundamental-mode resonance frequency versus *Q* factor for (a) 1LMoS$_2$ on 2LGr, (b) 1LMoS$_2$ on 3LGr, and (c) 1LMoS$_2$ on 4LGr hetero-structure nanomechanical resonators. The highest $f_0 \times Q$ is highlighted by purple dashed circles. Inset in each panel: Resonances data and its fitting curve for the device having the highest $f_0 \times Q$ in its category.

We now turn to discuss differences on resonance properties among graphene, MoS$_2$, and resulting heterostructures. The frequencies and *Q* factors of 2L graphene, 1L MoS$_2$ and resulting 1LMoS$_2$-2LGr are shown in **Figure 4**a and 4d. Generally, for devices of the same size (diameter), frequencies of heterostructure are higher than MoS$_2$ but lower than graphene. Similar trends are also observed in 1LMoS$_2$-3LGr (Fig. 4b) and 1LMoS$_2$-4LGr (Fig. 4c). By assuming the devices are in membrane regime, where the frequency is governed by built-in tension, the fundamental-mode resonance is given by

$$f_0 = \frac{2.404}{\pi d}\sqrt{\frac{\gamma}{\rho_{2D}}}, \qquad (1)$$

where *d* is device diameter, $\rho_{2D}$ is the areal mass density and $\gamma$ is built-in tension. For 2L graphene, 1L MoS$_2$ and 1LMoS$_2$-2LGr heterostructures devices, if we assume that graphene, MoS$_2$ and heterostructure have the same built-in tension level $\gamma$, since $\rho_{2D,2LGr} < \rho_{2D,1LMoS_2} < \rho_{2D,1LMoS_2\text{-}2LGr}$, we shall obtain $f_{2LGr} > f_{1LMoS_2} > f_{1LMoS_2\text{-}2LGr}$, which is not consistent well with our experimental results ($f_{2LGr} > f_{1LMoS_2\text{-}2LGr} > f_{1LMoS_2}$). This implies that the tension levels are different among the MoS$_2$, graphene, and the resulted heterostructures. Another interesting point is the *Q* factor difference among the graphene, MoS$_2$ and the heterostructured devices. It is observed that graphene devices exhibit higher $Q$ factors than those of their MoS$_2$ and heterostructure counterparts. Based on $Q$ factor equation $Q = f / \Gamma_m$, the $Q$ factor of graphene is indeed expected to be larger than that of MoS$_2$ (given the same diameter) assuming that $\Gamma_m$ (damping rate) remains the same for graphene and MoS$_2$ drumheads of the same diameter and thickness. However, though heterostructure devices exhibit higher frequencies than their MoS$_2$ counterparts, $Q$ factors of heterostructure devices are similar to, or even lower than, those of their MoS$_2$ counterparts. This suggests a possible increase of damping due to additional energy dissipation related to interlayer frictions at the MoS$_2$-graphene interface.

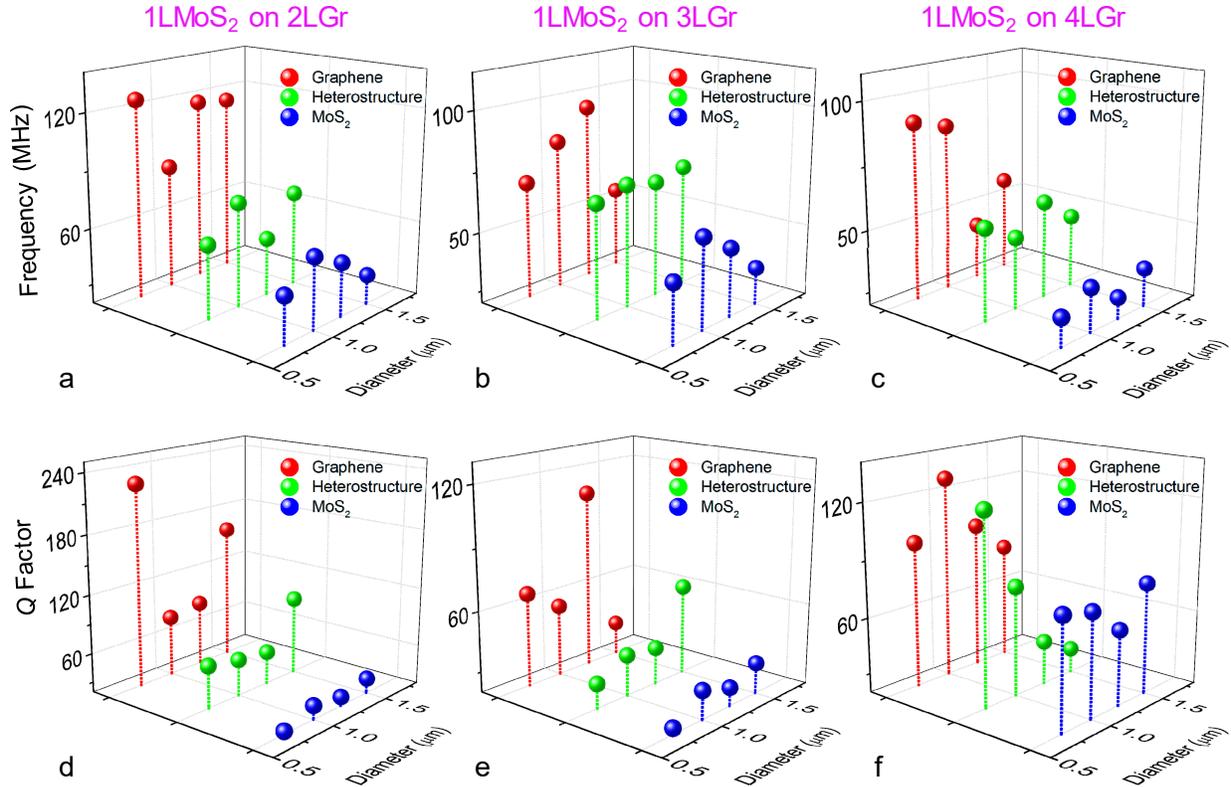

**Figure 4**: Comparison of the resonance frequencies measured from graphene, MoS$_2$ and their resulting heterostructures, for (a) 1L MoS$_2$ on 2L Gr, (b) 1L MoS$_2$ on 3L Gr, and (c) 1L MoS$_2$ on 4L Gr devices. Comparison of quality ($Q$) factors measured from graphene, MoS$_2$ and their resulting heterostructures in (d) 1L MoS$_2$ on 2L Gr, (e) 1L MoS$_2$ on 3L Gr, and (f) 1L MoS$_2$ on 4L Gr devices.

To further investigate the resonance frequency differences among the MoS$_2$, graphene, and the resulted heterostructures and probe their tension levels, analytical modeling is performed. The fundamental-mode resonance frequency of circular drumhead resonators can be expressed as [26,27]

$$f_0 = \left(\frac{k_m d}{4\pi}\right)\sqrt{\frac{16D}{\rho_{3D} t d^4}\left[\left(\frac{kd}{2}\right)^2 + \frac{\gamma d^2}{2D}\right]}, \qquad (2)$$

where $\rho_{3D}$ is the material mass density, $t$ is device thickness, $k_m$ is the modal parameter calculated numerically, $\gamma$ refers to the built-in tension and $D$ is the flexural rigidity, $D = E_Y t^3 / [12(1-\nu^2)]$, in which $E_Y$ and $\nu$ are Young's modulus and Poisson's ratio, respectively. As $\gamma d^2 / D \to \infty$, Eq. (2) approaches a membrane model, in which frequency is dominated by built-in tension. As $\gamma d^2 / D \to 0$, Eq. (2) approaches a plate model, in which frequency is dominated by Young's modulus $E_Y$. For heterostructures devices, Young's modulus and mass density vary as thickness changes. In our experiment, the MoS$_2$ thickness is fixed while graphene thickness increases from 1L to 4L, which gives us thickness of heterostructure as

$$t_{hetero} = t_{Gr} + 0.65 nm, \tag{3}$$

where 0.65nm is single-layer MoS$_2$ thickness. The Young's modulus of heterostructure is given by [28]

$$E_{Y,hetero} t_{hetero} = E_{Y,Gr} t_{Gr} + E_{Y,MoS_2} t_{Y,MoS_2}. \tag{4}$$

Combining Eq. (3) and Eq. (4), the Young's modulus of heterostructure is given by

$$E_{Y,hetero} = E_{Y,Gr} - (E_{Y,Gr} - E_{Y,MoS_2}) \frac{0.65 nm}{t_{Gr} + 0.65 nm}. \tag{5}$$

Similar, mass density of heterostructure is given by

$$\rho_{hetero} t_{hetero} = \rho_{Gr} t_{Gr} + \rho_{MoS_2} t_{MoS_2}, \tag{6}$$

and combining Eq. (4) and Eq. (6), the mass density of heterostructure is given by

$$\rho_{hetero} = \rho_{Gr} - (\rho_{Gr} - \rho_{MoS_2}) \frac{0.65 nm}{t_{Gr} + 0.65 nm}. \tag{7}$$

Combining Eq. (2), Eq. (5) and Eq. (7), the scaling curves of heterostructures with varying size are calculated and plotted, which are shown in **Figure 5**. To investigate how the frequency of heterostructure resonators is affected by MoS$_2$ and graphene, we also perform analytical model for MoS$_2$ and graphene using Eq. (3), also shown in Figure 5. We find that most devices in this work, including graphene, MoS$_2$ and heterostructures, are in the tension regime and transition regime, with tension levels between 0.05N/m to 0.5N/m. The frequency difference among graphene, MoS$_2$ and resulting heterostructures is mainly attributed to different tension levels: MoS$_2$ devices have lower tension levels compared with graphene and heterostructures, which is the main reason for MoS$_2$ resonators exhibiting the lowest frequency and heterostructures in the middle. Similar trends are also observed in devices of other sizes (Fig. 5b to 5d). This tension difference between MoS$_2$ and graphene could be explained by different sealing conditions caused by different fabrication methods. The graphene resonators are fabricated by direct exfoliation, which result in a relatively tight contact between flakes and substrates. As a result, air molecules trapped in the cavity cannot escape in a short term. When devices are preserved in vacuum conditions, air bulging effects lead to a higher tension levels hence higher fundamental-mode frequencies. In comparison, MoS$_2$ devices are fabricated using all dry-transfer method. Thus, the contact between MoS$_2$ flakes and substrates are not as tight as graphene flakes with substrates. As a result, air in cavity leak quickly when devices are in vacuum chamber, leading to lower tension levels and frequencies [29]. Based on the aforementioned discussions and experimental results, it can be concluded that resonance of heterostructure is dominated by bottom layer and affected by top layer, and an empirical equation describing the tension level could be expressed as,

$$\gamma_{\text{hetero}} = \gamma_{\text{top}} + \alpha \gamma_{\text{bottom}}, \tag{8}$$

where $\gamma_{\text{hetero}}$, $\gamma_{\text{top}}$ and $\gamma_{\text{bottom}}$ refer to the tension level of heterostructure, top layer and bottom layer respectively. The "interlayer coefficient" $\alpha$ describes how the tension level of heterostructure affected by bottom layer, which mainly depends on interface friction between top layer and bottom layer.

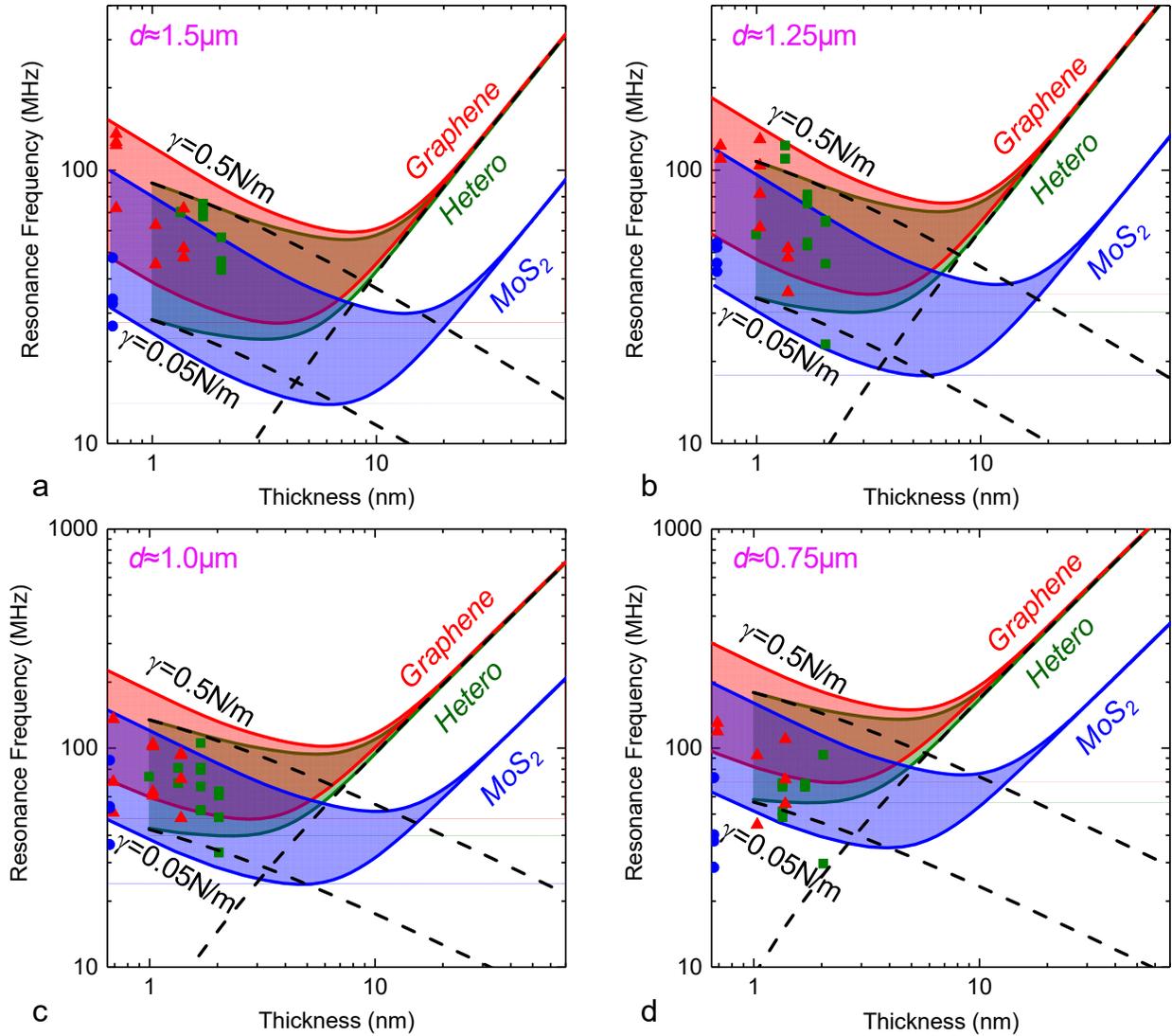

**Figure 5**: Frequency scaling of MoS$_2$-graphene van der Waals heterostructures for device diameters of (a) 1.5µm, (b) 1.25µm, (c) 1.0µm, and (d) 0.75µm. For each color, the upper solid line represents the calculated resonance frequency with a tension level of γ = 0.5N/m and the lower one of γ = 0.05N/m. The shadowed region shows the range of tension levels between 0.05N/m and 0.5N/m. Dashed lines illustrate the 'membrane' and 'disk' limits of elastic behavior. Squares, triangles, and circles (in colors consistent with those of the corresponding curves) represent measured data from heterostructure, graphene, and MoS$_2$ devices, respectively.

As the thickness of device increases, the frequency is determined by combination of Young's modulus and tension, which is a transition regime between tension model and plate model. Compared with MoS$_2$ and graphene, heterostructure devices enter the transition regime with even the smallest possible thickness. For diameter of 0.75μm and tension level of 0.05N/m (Fig. 5d), even the thinnest heterostructure devices, 1LGr-1LMoS$_2$, are already in transition regime, implying a depletion of tension regime. For heterostructure devices in plate regime, the frequencies are the same as those of the graphene devices. This is because, for thick devices, from Eq. (5), the Young's modulus of the heterostructures is mainly determined by the Young's modulus of graphene.

## Conclusions

In conclusion, we have demonstrated, for the first time, freestanding atomic layer MoS$_2$-on-graphene van der Waals heterostructures, based on which we have further realized nanomechanical resonators. All heterostructure devices exhibit robust resonances up to ~100 MHz in the VHF band, with figure-of-merit as high as $f_0 \times Q \approx 8.7 \times 10^9$ Hz. We observe high uniformity in the resonance frequencies and $Q$ factors measured from these heterostructure resonators. The resonance frequencies and $Q$ factors of the heterostructure devices are found to be in the middle (a compromise) between those of the devices based on a single constituting crystal (graphene or MoS$_2$). The measurement and analysis suggest that interlayer interactions play an important role in setting the tension level and damping of heterostructured resonators. This study not only opens new opportunities for studying multiphysical coupling effects in hetero-atomic-layer and bimorph 2D devices, but also sheds light on engineering mechanical degrees of freedom and interlayer interactions in van der Waals heterostructures.

# Methods

**Suspended MoS$_2$-Graphene Heterostructure Device Fabrication**

The fabrication of MoS$_2$-graphene heterostructures and prototype bimorph devices starts from exfoliating graphene on SiO$_2$-on-Si substrates with pre-patterned arrays of microtrenches. After obtaining atomically thin graphene, single-layer (1L) MoS$_2$ flakes are carefully selected and transferred on top of the graphene flakes, by using an all-dry transfer method with alignment [22]. CVD MoS$_2$ layers are used in 1LMoS$_2$ on 1Lgraphene (Gr), 1LMoS$_2$ on 2LGr, and 1LMoS$_2$ on 3LGr devices, while exfoliated MoS$_2$ flakes are used in making 1LMoS$_2$ on 4LGr devices.

**Raman Scattering Measurement**

Heterostructure devices are preserved in a vacuum chamber and measured using customized micro-Raman system that is integrated into an optical interferometric resonance measurement system (please see the section below). The 532nm laser is focused on the center of the heterostructure device with a typical spot size around ~1µm and laser power below 200µW, to avoid excessive laser heating. Raman scattered light from the sample is collected in backscattering geometry and then guided to a spectrometer (Horiba iHR550) with a grating of 2400 g per mm. Raman signal is recorded using a liquid-nitrogen-cooled CCD. The spectral resolution of this system is ~1cm$^{-1}$.

**Interferometric Resonance Measurement**

We study MoS$_2$-graphene heterostructure devices and their nanomechanical resonances using an ultrasensitive laser interferometry system (Fig. 1b). The resonant motion is photothermally excited by using an amplitude modulated 405nm blue laser. To avoid excessive laser heating, laser is focused ~5µm away from the devices with laser power below 300µW. A network analyzer is used to control the modulation depth and frequency for modulating the 405nm laser, sweeping from 1MHz to 150MHz. The resonance motion is detected by a 633nm red laser with average power of 600µW. Typical laser spot sizes are ~5µm and ~1µm for the 405nm and 633nm lasers, respectively. The output signal in the frequency domain is recorded by the same network analyzer.

**Acknowledgement**: We thank the financial support from National Science Foundation CAREER Award (Grant ECCS-1454570) and CCSS Award (Grant ECCS-1509721). Part of the device fabrication was performed at the Cornell NanoScale Science and Technology Facility (CNF), a member of the National Nanotechnology Infrastructure Network (NNIN), supported by the National Science Foundation (Grant ECCS-0335765).

*– Supplementary Information –*

# Atomic Layer MoS$_2$-Graphene van der Waals Heterostructure Nanomechanical Resonators

Fan Ye, Jaesung Lee, Philip X.-L. Feng[*]

*Department of Electrical Engineering & Computer Science, Case School of Engineering, Case Western Reserve University, 10900 Euclid Avenue, Cleveland, OH 44106, USA*

**Table S1**: Frequency and Quality ($Q$) Factors of Atomic Layer Graphene, MoS$_2$ and Their Vertically Stacked Heterostructure Nanomechanical Resonators

**1L MoS$_2$ / 2L Graphene Heterostructures**

**Diameter $d \approx 0.75$μm**

| Device # | 2L Graphene | | Device # | 1L MoS$_2$ (CVD) | | Device # | Heterostructure | |
|---|---|---|---|---|---|---|---|---|
| | $f$ (MHz) | $Q$ Factor | | $f$ (MHz) | $Q$ Factor | | $f$ (MHz) | $Q$ Factor |
| 1 | 119.1 | 181 | 1 | 28.5 | 18 | 1 | 51.4 | 30 |
| 2 | 130.2 | 270 | 2 | 40.2 | 18 | 2 | 48.4 | 18 |
| | | | 3 | 73.3 | 33 | 3 | 69.2 | 125 |
| | | | 4 | 37.9 | 24 | 4 | 66.7 | 77 |
| Normal Distributed Mean Value | 124.6±7.8 | 226±63 | | 45.0±19.5 | 23±7 | | 58.9±10.5 | 63±48 |

**Diameter $d \approx 1.00$μm**

| Device # | 2L Graphene | | Device # | 1L MoS$_2$ (CVD) | | Device # | Heterostructure | |
|---|---|---|---|---|---|---|---|---|
| | $f$ (MHz) | $Q$ Factor | | $f$ (MHz) | $Q$ Factor | | $f$ (MHz) | $Q$ Factor |
| 1 | 50.7 | 22 | 1 | 36.3 | 28 | 1 | 69.2 | 39 |
| 2 | 135.8 | 123 | 2 | 88.0 | 19 | 2 | 81.2 | 75 |
| 3 | 70.2 | 97 | 3 | 54.1 | 39 | | | |
| | | | 4 | 53.2 | 51 | | | |
| Normal Distributed Mean Value | 85.4±44.8 | 81±51 | | 57.9±51.7 | 34±14 | | 75.2±8.5 | 57±25 |

[*]Corresponding Author. Email: philip.feng@case.edu

## Diameter $d \approx 1.25\mu m$

| Device # | 2L Graphene | | Device # | 1L MoS$_2$ (CVD) | | Device # | Heterostructure | |
|---|---|---|---|---|---|---|---|---|
| | $f$ (MHz) | $Q$ Factor | | $f$ (MHz) | $Q$ Factor | | $f$ (MHz) | $Q$ Factor |
| 1 | 123.0 | 89 | 1 | 45.7 | 30 | 1 | 31.6 | 13 |
| 2 | 110.1 | 82 | 2 | 54.2 | 26 | 2 | 69.7 | 94 |
| | | | 3 | 52.2 | 22 | | | |
| | | | 4 | 42.6 | 39 | | | |
| Normal Distributed Mean Value | 116.6±9.1 | 86±5 | | 48.7±5.4 | 29±8 | | 50.6±26.8 | 53±57 |

## Diameter $d \approx 1.50\mu m$

| Device # | 2L Graphene | | Device # | 1L MoS$_2$ (CVD) | | Device # | Heterostructure | |
|---|---|---|---|---|---|---|---|---|
| | $f$ (MHz) | $Q$ Factor | | $f$ (MHz) | $Q$ Factor | | $f$ (MHz) | $Q$ Factor |
| 1 | 135.6 | 185 | 1 | 47.8 | 24 | 1 | 70.4 | 95 |
| 2 | 126.7 | 173 | 2 | 26.9 | 29 | 2 | 70.3 | 106 |
| 3 | 72.6 | 89 | 3 | 33.8 | 44 | 3 | 70.2 | 96 |
| 4 | 123.3 | 183 | 4 | 32.5 | 44 | | | |
| Normal Distributed Mean Value | 114.6±28.4 | 158±46 | | 35.3±9.3 | 35±10 | | 70.3±0.1 | 99±6 |

## 1L MoS$_2$ / 3L Graphene Heterostructures

### Diameter $d \approx 0.75\mu m$

| Device # | 3L Graphene | | Device # | 1L MoS$_2$ (CVD) | | Device # | Heterostructure | |
|---|---|---|---|---|---|---|---|---|
| | $f$ (MHz) | $Q$ Factor | | $f$ (MHz) | $Q$ Factor | | $f$ (MHz) | $Q$ Factor |
| 1 | 44.7 | 16 | 1 | 28.5 | 18 | 1 | 66.5 | 31 |
| 2 | 92.6 | 116 | 2 | 40.2 | 18 | 2 | 68.6 | 33 |
| | | | 3 | 73.3 | 33 | | | |
| | | | 4 | 37.9 | 24 | | | |
| Normal Distributed Mean Value | 68.7±33.9 | 66±70 | | 45.0±19.5 | 23±7 | | 67.6±1.5 | 32±1 |

## Diameter $d \approx 1.00\,\mu m$

| Device # | 3L Graphene | | Device # | 1L MoS$_2$ (CVD) | | Device # | Heterostructure | |
|---|---|---|---|---|---|---|---|---|
| | $f$ (MHz) | $Q$ Factor | | $f$ (MHz) | $Q$ Factor | | $f$ (MHz) | $Q$ Factor |
| 1 | 60.6 | 12 | 1 | 36.3 | 28 | 1 | 52.0 | 47 |
| 2 | 102.0 | 29 | 2 | 88.0 | 19 | 2 | 52.2 | 29 |
| 3 | 63.1 | 52 | 3 | 54.1 | 39 | 3 | 105.2 | 62 |
| 4 | 104.2 | 127 | 4 | 53.2 | 51 | 4 | 66.8 | 30 |
| | | | | | | 5 | 79.5 | 32 |
| Normal Distributed Mean Value | 82.5±23.9 | 55±51 | | 57.9±51.7 | 34±14 | | 71.1±22.2 | 40±14 |

## Diameter $d \approx 1.25\,\mu m$

| Device # | 3L Graphene | | Device # | 1L MoS$_2$ (CVD) | | Device # | Heterostructure | |
|---|---|---|---|---|---|---|---|---|
| | $f$ (MHz) | $Q$ Factor | | $f$ (MHz) | $Q$ Factor | | $f$ (MHz) | $Q$ Factor |
| 1 | 103.9 | 91 | 1 | 45.7 | 30 | 1 | 81.4 | 84 |
| 2 | 81.8 | 56 | 2 | 54.2 | 26 | 2 | 52.9 | 28 |
| 3 | 61.8 | 67 | 3 | 52.2 | 22 | 3 | 75.5 | 23 |
| 4 | 129.7 | 197 | 4 | 42.6 | 39 | 4 | 78.7 | 27 |
| | | | | | | 5 | 54.3 | 28 |
| Normal Distributed Mean Value | 94.3±29.2 | 103±65 | | 48.7±5.4 | 29±8 | | 68.4±14.1 | 38±26 |

## Diameter $d \approx 1.50\,\mu m$

| Device # | 3L Graphene | | Device # | 1L MoS$_2$ (CVD) | | Device # | Heterostructure | |
|---|---|---|---|---|---|---|---|---|
| | $f$ (MHz) | $Q$ Factor | | $f$ (MHz) | $Q$ Factor | | $f$ (MHz) | $Q$ Factor |
| 1 | 45.3 | 19 | 1 | 47.8 | 24 | 1 | 67.6 | 83 |
| 2 | 63.0 | 53 | 2 | 26.9 | 29 | 2 | 70.6 | 79 |
| | | | 3 | 33.8 | 44 | 3 | 75.4 | 30 |
| | | | 4 | 32.5 | 44 | | | |
| Normal Distributed Mean Value | 54.2±12.5 | 36±24 | | 35.3±9.3 | 35±10 | | 71.2±3.9 | 64±30 |

## 1L MoS$_2$ / 4L Graphene Heterostructures

### Diameter $d \approx 0.75 \mu m$

| Device # | 4L Graphene | | Device # | 1L MoS$_2$ (Exfoliated) | | Device # | Heterostructure | |
|---|---|---|---|---|---|---|---|---|
| | $f$ (MHz) | $Q$ Factor | | $f$ (MHz) | $Q$ Factor | | $f$ (MHz) | $Q$ Factor |
| 1 | 71.9 | 76 | 1 | 24.2 | 23 | 1 | 29.7 | 215 |
| 2 | 109.8 | 118 | 2 | 43.7 | 113 | 2 | 93.2 | 27 |
| | | | 3 | 14.2 | 9 | | | |
| | | | 4 | 27.6 | 76 | | | |
| | | | 5 | 47.2 | 174 | | | |
| Normal Distributed Mean Value | 90.5±26.2 | 97±30 | | 31.4±13.8 | 79±67 | | 61.5±44.9 | 121±96 |

### Diameter $d \approx 1.00 \mu m$

| Device # | 4L Graphene | | Device # | 1L MoS$_2$ (Exfoliated) | | Device # | Heterostructure | |
|---|---|---|---|---|---|---|---|---|
| | $f$ (MHz) | $Q$ Factor | | $f$ (MHz) | $Q$ Factor | | $f$ (MHz) | $Q$ Factor |
| 1 | 72.3 | 45 | 1 | 33.6 | 104 | 1 | 33.4 | 122 |
| 2 | 92.8 | 132 | 2 | 42.5 | 95 | 2 | 63.4 | 32 |
| 3 | 93.9 | 203 | 3 | 31.4 | 37 | | | |
| | | | 4 | 31.9 | 31 | | | |
| | | | 5 | 29.6 | 35 | | | |
| | | | 6 | 46.5 | 71 | | | |
| | | | 7 | 43.9 | 154 | | | |
| Normal Distributed Mean Value | 86.3±12.2 | 127±79 | | 37.5±7.0 | 75±46 | | 48.4±21.2 | 77±64 |

### Diameter $d \approx 1.25\,\mu m$

| Device # | 4L Graphene | | Device # | 1L MoS$_2$ (Exfoliated) | | Device # | Heterostructure | |
|---|---|---|---|---|---|---|---|---|
| | $f$ (MHz) | $Q$ Factor | | $f$ (MHz) | $Q$ Factor | | $f$ (MHz) | $Q$ Factor |
| 1 | 35.8 | 122 | 1 | 17.1 | 114 | 1 | 45.5 | 69 |
| 2 | 47.9 | 74 | 2 | 25.3 | 52 | 2 | 65.4 | 30 |
| | | | 3 | 26.9 | 21 | 3 | 64.7 | 39 |
| | | | 4 | 26.6 | 44 | 4 | 23.1 | 18 |
| | | | 5 | 35.1 | 51 | | | |
| | | | 6 | 34.5 | 137 | | | |
| | | | 7 | 7.55 | 8 | | | |
| | | | 8 | 35.5 | 157 | | | |
| Normal Distributed Mean Value | 41.9±8.6 | 98±35 | | 26.1±9.8 | 73±56 | | 49.7±20.0 | 39±22 |

### Diameter $d \approx 1.50\,\mu m$

| Device # | 4L Graphene | | Device # | 1L MoS$_2$ (Exfoliated) | | Device # | Heterostructure | |
|---|---|---|---|---|---|---|---|---|
| | $f$ (MHz) | $Q$ Factor | | $f$ (MHz) | $Q$ Factor | | $f$ (MHz) | $Q$ Factor |
| 1 | 47.9 | 132 | 1 | 18.8 | 81 | 1 | 43.1 | 32 |
| 2 | 72.4 | 52 | 2 | 23.9 | 25 | 2 | 46.5 | 30 |
| 3 | 51.8 | 61 | 3 | 51.2 | 68 | 3 | 56.8 | 36 |
| | | | 4 | 23.2 | 82 | | | |
| | | | 5 | 66.1 | 16 | | | |
| | | | 6 | 28.9 | 203 | | | |
| Normal Distributed Mean Value | 57.4±13.2 | 82±44 | | 35.3±18.9 | 79±67 | | 48.8±7.1 | 33±3 |